# Astro2020 Project White Paper

# The Cosmic Accelerometer

**Thematic Areas:**  ☒ Planetary Systems  ☐ Star and Planet Formation
☐ Formation and Evolution of Compact Objects  ☒ Cosmology and Fundamental Physics
☐ Stars and Stellar Evolution  ☐ Resolved Stellar Populations and their Environments
☐ Galaxy Evolution  ☐ Multi-Messenger Astronomy and Astrophysics

**Principal Authors:**
Name: Stephen Eikenberry & Anthony Gonzalez
Institution: University of Florida
Email:  eiken@ufl.edu, anthonyhg@ufl.edu
Phone: 352-294-1833

**Co-authors:** (names and institutions)
Jeremy Darling (University of Colorado), Jochen Liske (Hamburger Sternwarte), Zachary Slepian (University of Florida), Guido Mueller (University of Florida), John Conklin (University of Florida), Paul Fulda (University of Florida), Claudia Mendes de Oliveira (Universidade de Sao Paulo), Misty Bentz (Georgia State University), Sarik Jeram (University of Florida), Chenxing Dong (University of Florida), Amanda Townsend (University of Florida), Lilianne Mariko Izuti Nakazono (Universidade de Sao Paulo), Robert Quimby (San Diego State University), William Welsh (San Diego State University), Joseph Harrington (University of Central Florida), Nicholas Law (University of North Carolina – Chapel Hill)

**Abstract**: We propose an experiment, the Cosmic Accelerometer, designed to yield velocity precision of ≤1 cm/s with measurement stability over years to decades. The first-phase Cosmic Accelerometer, which is at the scale of the Astro2020 "Small" programs, will be ideal for precision radial velocity measurements of terrestrial exoplanets in the Habitable Zone of Sun-like stars. At the same time, this experiment will serve as the technical pathfinder and facility core for a second-phase larger facility at the "Medium" scale, which can provide a significant detection of cosmological redshift drift on a 6-year timescale. This larger facility will naturally provide further detection/study of "Earth twin" planet systems as part of its external calibration process. This experiment is fundamentally enabled by a novel low-cost telescope technology called PolyOculus, which harnesses recent advances in commercial off the shelf equipment (telescopes, CCD cameras, and control computers) combined with a novel optical architecture to produce telescope collecting areas equivalent to standard telescopes with large mirror diameters. Combining a PolyOculus array with an actively-stabilized high-precision radial velocity spectrograph provides a unique facility with novel calibration features to achieve the performance requirements for the Cosmic Accelerometer.

# I. Key Goals and Objectives

Several of the key science areas of the coming decade, from exoplanets and astrobiology to dark matter and dark energy, can be addressed using precision velocity measurements that yield acceleration determinations on month to decade timescales. Such measurements will be critical to identifying and determining the properties of Earth-mass planets around Sun-like stars — a key science area for the coming decade and a key target space for related searches for life in the Universe (see Astro2020 Science White Papers [1-7]). In addition, the detection of cosmological redshift drift – the change in the cosmological expansion rate with time — can provide direct detection of the accelerating expansion of the Universe due to dark energy (see Astro2020 Science White Papers [8,9]). However, the latter requires velocity measurements of ≤1 cm/s over timescales of years to decades [10]. Similarly, habitable planets around Sun-like stars have Earth-like orbits, with reflex velocities of <10 cm/s and periods of ~1 yr. Long-term (years) timescale stability is thus also a crucial ingredient for discovery and study of these planets: without this, radial velocity studies for mass determination simply cannot be done.

We describe below an experiment, the Cosmic Accelerometer, designed to yield velocity precision of ≤1 cm/s, with measurement stability over years to decades. The first-phase Cosmic Accelerometer, which we will develop and deploy as part of the "Small" programs in the Astro2020 timeline, would be ideal for precision radial velocity measurements of terrestrial exoplanets in the Habitable Zone of Sun-like stars. At the same time, this experiment will serve as the technical pathfinder and facility core for a second-phase larger facility at the "Medium" scale, which can provide a significant detection of cosmological redshift drift on a 6-year timescale. This larger facility will naturally provide further detection/study of "Earth twin" planet systems as part of its external calibration process. This facility will also be able to constrain the evolution of fundamental constants and potentially observe accelerations of hypervelocity stars and galaxies in nearby clusters.

# II. Technical Overview: Cosmic Accelerometer Design

## A. Experimental Design Overview and Key Drivers

In the following discussion, we will focus on the design drivers for the cosmological redshift drift aspect of the proposed overall experiment. An experiment optimized for redshift drift also achieves the performance required for the exoplanet science to be carried out by the proposed Phase I Cosmic Accelerometer.

Eikenberry & Gonzalez et al. [8] describes the key drivers for this experiment in greater detail, and we provide a brief summary here. Detection of redshift drift [11,12] requires redshift measurements of spectral features that, at least in an ensemble average, are stable over the period of the experiment and have relatively low intrinsic linewidths. Use of the Lyα forest to measure redshift drift was proposed by Loeb [13]. The idea in this case is to use quasars as backlights and measure the drift of the Lyα forest lines. While this approach is restricted to redshifts at which Lyα is observable, a single observation will contain many lines that can be used to precisely infer d$z$/d$t$ as a function of redshift over the observable window. Other groups [9] have suggested that [OII] doublets may provide an alternative or complement to the Lyα forest for measuring redshift drift. While secular changes in the [OII] line profiles will dominate the measurement errors, a differential measurement of the doublet line spacing should be independent of that effect. In addition, these features are observable from the ground at z<1, probing the epoch when dark energy is dominant. The Cosmic Accelerometer can potentially make these measurements as well, though this is not part of the baseline strategy.

The fundamental challenge for measuring cosmological redshift drift is that the signal is small. Detection requires measuring redshifts with an accuracy of a few parts in $10^{10}$, or equivalently being able to detect velocity changes at the level of ~1 cm/s/yr, with an instrument that is stable at this precision over the timescale of the measurement. Liske et al. [10] provided a detailed analysis of the feasibility of redshift drift detection using ELT-class collecting areas, concluding that significant detections would be possible on multi-decade timescales. This motivated a design study for the COsmic Dynamics and EXo-earth experiment (CODEX) for the European Extremely Large Telescope (E-ELT), a precursor to the currently-planned second-generation instrument HIRES. Redshift drift is one of the key science cases for this instrument, which has completed its Phase A study and is forecast to be on sky by the late 2020s. Given this timescale, a detection of redshift drift by 2050 is plausible. While this observatory-class facility has the potential to detect cosmic acceleration by 2050, the US community does not have access to it, and the timeline of 30 years is untenable given currently pressing questions on the expansion history of the Universe and the role of dark energy. **For these reasons, we advocate a dedicated cosmic dynamics experiment that is both more cost-effective and will deliver results on much shorter timescales than are achievable with these large facilities.**

The Cosmic Accelerometer can achieve the requirements outlined above by addressing the 3 major challenges they present. The first (Key Requirement 1) is the faintness of the targets compared to the extreme precision required, demanding very large telescope collecting area [10]. The Cosmic Accelerometer addresses this via the novel PolyOculus technology, delivering a large-aperture-equivalent telescope using fiber optics to link arrays of semi-autonomous, small, inexpensive telescopes — enabling effective areas 10x larger at the same cost. The second challenge (Key Requirement 2) is the extreme precision (≤1 cm/s) velocity measurements required [8,9]. We address this with an actively-stabilized high-resolution spectrograph that offers outstanding velocity stability via ultra-precision interferometric sensing/control techniques. The third critical challenge (Key Requirement 3) is maintaining this velocity precision on timescales of >10 years [8,9]. For this problem, we combine the unique features of the PolyOculus telescope array with the stabilized spectrograph to derive a novel calibration of the system and hold velocity drifts to <0.1 cm/s/year based on stellar ensemble velocities.

Below we describe the key features of the Cosmic Accelerometer, including the telescope or observatory architecture, key performance requirements, and technical requirements.

### B. Key Requirement 1: Large Collecting Area – the PolyOculus Telescope Array

The first key requirement for the Cosmic Accelerometer is directly tied to the telescope/observatory architecture for it. As discussed by Liske et al. [10], use of the Ly$\alpha$ forest to measure redshift drift requires a telescope aperture on the scale of an ELT-class telescope — though not the diffraction-limited performance. In the subsections below, we address first the required collecting area and then the possible solution to achieving such an area at relatively low cost.

*Required Telescope Collecting Area:* We follow the approach of Liske et al. [10] to determine the size of the collecting area needed, verifying their conclusion that an ELT-class telescope used for a fraction of its observing time could measure redshift drift on 20-year timescales. We also run detailed numerical simulations of the redshift drift measurement, using the dedicated experiment and the stabilization/calibration approach outlined above. For these simulations, we use actual high-resolution Lyα spectra of the QSO SDSS J121843.39+153617.2 from archival Keck HIRES spectrograph observations [14,15] to determine a typical distribution of Lyα line parameters, including HI column densities and line width parameters. We create an initial spectrum matching the expected number of lines per wavelength range, and then add realistic noise effects (photon shot noise, dark current, read noise, etc.) to obtain a simulated observed spectrum. We find that for the expected parameters (see below for system details), the measurement noise will be dominated by target shot noise effects (assuming a stable HPRV spectrograph at the 1 cm/s level). We next use a template redshifted spectrum to extract a measured redshift via cross-correlation. We repeated this for a large number (~$10^3$) of simulated noise realizations, and then extracted the RMS of the resulting redshift distributions as our measurement noise. We carried out a trade study for various combinations of telescope area, observation time, and cost, identifying an optimal solution in terms of cost and ability to distinguish models of dark energy. We present in Figure 1 the expected results after a 6-year observing campaign on a 15m-equivalent telescope. We find that this approach yields a >3σ detection of cosmological acceleration, as expected from the analytical estimate above. **This will be the first direct detection of cosmological acceleration, and provide an important confirmation of fundamental cosmological principles**.

The signal-to-noise ratio for redshift drift measurements grows as ~$t^{3/2}$ — a factor of $t^{1/2}$ due to the improved signal-to-noise on the redshifts themselves, and $t^1$ due to the increased time baseline. Thus, extended operation of the same facility enables constraints on the nature of dark energy. In Figure 1 we also show simulated results for a 10-year operational observing run on a 15m-equivalent facility. Some classes of dark energy models will be ruled out by such observations, and many others strongly constrained. Being a dynamical rather than geometric test – and one that is physically simple — redshift drift provides a unique and independent means of confirming the existence of dark energy and validating conclusions from other probes.

As we show below, a unique approach can make this aperture equivalent affordable and highly scientifically productive not only for the redshift drift experiment, but also for studies of exoplanets and evolution of fundamental physical constants [16].

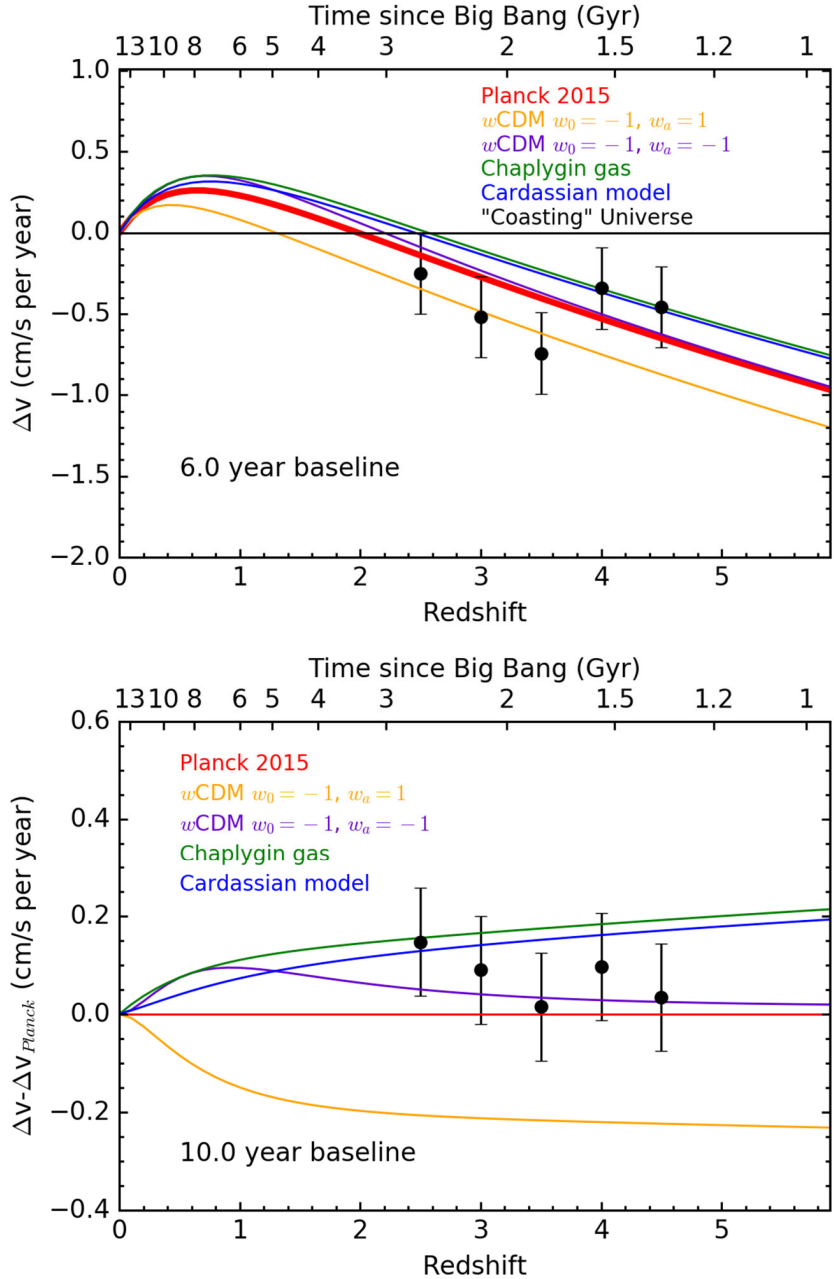

*Figure 1* – *(Top) Expected redshift drift versus redshift over a 6-year measurement baseline with a 15m-equvalent collecting aperture. Note that a "coasting Universe" can be easily distinguished from models containing dark matter/energy with a precision of ~1 cm/s. The displayed data points provide a p<1.5x10$^{-3}$ confidence interval (equivalent to 3.2σ) for the detection of acceleration. Using the Liske formalism [10] as an independent cross-check, we estimate a 3σ detection in ~7 yr. (Bottom)* – *Expected redshift drift versus redshift over a 10-year measurement baseline with a 15m-equivalent collecting aperture. (This is a different statistical realization of the simulated data from the top frame). For both plots, we plot a standard LCDM cosmology [17], models with evolving w, and two alternative cosmologies for illustration — a Chaplygin gas and Cardassian models. For the latter we use the best fit constraints derived from current data sets presented in [18] and [19], respectively. In both plots, we assume a standard LCDM cosmology to generate the simulated data points.*

*The PolyOculus Telescope Array:* The PolyOculus approach matches the power of large-aperture telescopes by using fiber optics to link modules of commercial-off-the-shelf (COTS) telescopes. These telescopes are semi-autonomous, small, and inexpensive. Crucially, this scalable design has construction costs 10x lower than equivalent traditional large-area telescopes. This approach relies primarily on recent advances in low-cost, high-performance COTS equipment — telescopes, CCD cameras, and control computers — combined with a novel fiber-link architecture and a few key technological innovations to produce telescope collecting areas equivalent to standard telescopes with mirror diameters ranging from 0.9-m to tens of meters (and beyond) for certain applications. The basic PolyOculus module in this scalable design consists of a "7-pack" of small COTS telescopes which are semi-autonomous in their pointing, tracking, and focus control, and equipped with a novel autonomous acquisition/guiding system and a fiber optic feed. The fiber optic outputs are linked together via a novel switchable optical coupling system which also provides intensity metering for the automated acquisition system. The linked 7-pack output can then be further linked to other 7-packs into a hierarchical, scalable architecture, or it can be fed directly into a common, shared detection system. The switchable coupling approach allows the dynamic reconfiguration of this system on timescales of ~60 seconds from a single large aperture to multiple, independently-pointable/operable smaller apertures.

One of the fundamental benefits of the PolyOculus approach is its extremely low-cost combined with high performance. Based on our calculations and simulations, and confirmed with our own laboratory tests and published results on fiber coupling efficiency from other groups, the net performance of this approach for a 2.5m-equivalent aperture (for a 7x 7-pack PolyOculus) is similar to that of a "standard" 2.5m telescope with an additional throughput loss of ~15% (sensitivity reduction of ~7-8%). Meanwhile, the total cost per square meter of such a PolyOculus system is under $200,000. For comparison, recent commercial quotes for a standard telescope from established vendors, including enclosure and guiding system range from $1,000,000 to $1,500,000 per square meter. In other words, with a slight reduction in sensitivity, PolyOculus can reduce costs for telescope apertures to ~20% of the cost of standard telescopes – and for larger apertures this can reduce to ~10% due to an economy of scale (see Figure 2).

### C. Key Requirement 2: Precision Radial Velocities - An Actively Stabilized Spectrograph

As shown in the previous section, the experiment must also have a spectrograph with ≤1 cm/s precision and stability on timescales from years to decades to achieve the science goals for cosmological acceleration. Earth-mass planets in the Habitable Zone of Sun-like stars will have projected stellar reflex velocities of ≤ 10 cm/s on timescales of ~1 year – implying similar requirements for that science area as well. Standard spectrographs for exoplanet radial velocity measurements have in recent years achieved regular performance of ~10 cm/s on timescales of days or longer, and have even reached ~3 cm/s through the use of Laser Frequency Comb (LFC) calibrations – albeit with stability timescales of hours or days [20]. The LFC systems can be internally stable to <0.1 cm/s on timescales of decades – the limiting factor here is that the spectrographs drift in velocity space, even on the timescales of the calibrations/observations themselves. Thus, achieving the performance requirements for the science areas we consider here requires a spectrograph with much greater stability than currently available.

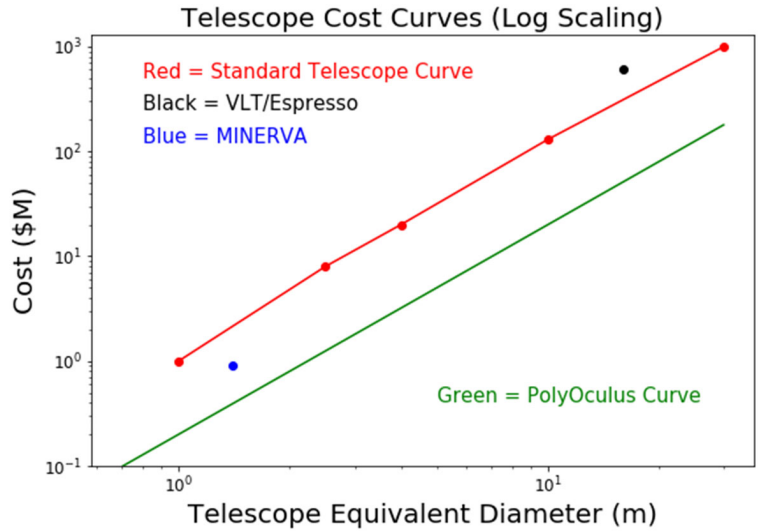

*Figure 2* – *Comparison of telescope costs (including enclosures, but no instruments) for standard telescopes versus PolyOculus arrays. PolyOculus arrays are substantially (5-10x) less expensive than standard telescopes (prices are current-year dollars, drawn from various sources via private communication). We note that MINERVA [21] also provides imaging, so this is not a completely "fair" comparison — but it is valid for spectroscopy-only applications.*

Current high-precision radial velocity (HPRV) spectrographs are effectively passively controlled systems [22]. While there may be temperature stabilization of the spectrograph to minimize thermal expansion, there is no active monitoring/correction of path length changes within the spectrographs, which lead to drift in the observed radial velocity measurements. The goal of ≤1 cm/s stability translates to dimensional stability of ~5nm for typical spectrograph resolutions and layouts considered here – a goal which is still beyond standard approaches.

We thus turn to the example of gravitational wave detectors for guidance. Team member G. Mueller led the laser input photonics for the Laser Interferometer Gravitational-wave Observatory (LIGO), which used active laser control systems to achieve stability orders of magnitude better than needed for our application here. Mueller and team members P. Fulda and J. Conklin are also leading US participation in the LISA mission for a space gravitational wave observatory, and Conklin also participated in the successful LISA Pathfinder mission. In all of these cases, the most sensitive detectors use active control as well as passive isolation to achieve their precision stability goals – we propose a similar approach here to an actively-stabilized HPRV spectrograph.

The spectrograph will employ cryogenic, vacuum, and vibration isolation technologies based on successful team experience with previous astronomical instruments. The optical bench will also be mounted with LISA-style interferometric Nd:YAG laser cavities that provide metrology of the key optical dimensions to an accuracy of ~0.1 nm RMS on timescales of minutes (the typical drift timescales for well-isolated spectrographs). This metrology will then be fed into an active control loop using thermal heating pads bonded to the bench to provide dimensional control at the level of ≤1 nm. We will thus effectively "steer" the bench shape to be stable to the required precision over the timescale needed for LFC calibration.

This approach provides an HPRV spectrograph with the required precision of ≤1 cm/s over timescales from seconds to years. However, while the actively-stabilized spectrograph approach is a necessary step in the direction needed, it is critical that the velocity precision of the system is robust to changes associated with instrument and telescope maintenance on multi-year timescales. For this reason, a central element of the experimental design is inclusion of external calibration sources selected to yield a null reference signal. As we show in the next section, **the combination of an actively-controlled HPRV spectrograph with the unique features of the PolyOculus telescope array enables such an external calibration to confirm the stability of the system on decade timescales.**

### D. Key Requirement 3: Long-term Stability – Stellar Ensemble Calibration

A key feature of the PolyOculus technology is its ability to separate out sub-arrays to point them at other targets on the sky with wide angle separations, and feed those into the HPRV spectrograph in parallel to the primary (QSO) targets. This reconfiguration and pointing takes place on a timescale of <5 minutes. This enables a unique observing strategy, whereby the majority of the telescope apertures are observing the QSO target for direct cosmological redshift measurements. At the same time, several smaller sub-arrays can be pointed at much brighter stars in the sky, making HPRV measurements of them. Even "quiet" stars have intrinsic radial velocity noise due to astrophysical processes, with amplitudes of a few x 10 cm/s [23].

Furthermore, in addition to the known short-term jitter at these amplitudes, the stars may also have secular RV trends that last years or decades. However, the ensemble average velocity of many stars should be very close to zero after barycentric corrections. Thus, by observing multiple stars (with typical observations of <30 minutes for bright RV-quiet stars in a few sub-arrays simultaneously) in parallel with the QSO observations, we can build up an ensemble of >$10^3$ stellar RVs on timescales of a few weeks. Even for stellar velocity noise amplitudes of ~30 cm/s, we can thus achieve an external reference which is stable to <1 cm/s. Furthermore, by cycling through different sub-arrays – so that each sub-array participates in the calibration process – we can effectively eliminate non-common-path errors between the different science sub-arrays. This then provides a "stellar ensemble calibration" which in conjunction with the LFC internal calibration extends the stability of the radial velocities to the required decades-long timescales, all while eliminating non-common-path errors.

Selecting RV-quiet reference stars will require some advance observational work and planning. The primary source of "contamination" for these stars will be previously-unknown low-mass planets in long-period (>months) orbits. However, the analysis of the calibration data should allow us to identify these systems as outliers and remove them from the calibration database over time. At the same time, low-mass planets in years-long orbits are exactly the primary targets for searches for extraterrestrial life described in Section I. Thus, a key "bycatch" product of the redshift drift experiment will be Earth-mass planets in the Habitable Zones of Sun-like stars.

### E. The Phase I Cosmic Accelerometer

As described above, on a five-year timescale the plan is deployment and operation of a first-phase Cosmic Accelerometer focusing on precision radial velocity measurements of terrestrial exoplanets in the Habitable Zone of Sun-like stars. At the same time, this will serve as the technical pathfinder and core array for a second-phase larger facility capable of providing a significant detection of cosmological redshift drift on a <5-year timescale. The sections above show how a PolyOculus telescope array coupled to an actively-stabilized HPRV spectrograph with a Laser Frequency Comb and using the stellar ensemble observing/calibration strategy can provide ≤1 cm/s velocity precision over years-long timescales. These are exactly the performance requirements for detection and mass measurement of Earth-mass planets in Habitable Zone orbits around Sun-like stars, which are the science targets for the Phase I Cosmic Accelerometer. In the subsections below, we describe the technical approach for this.

*PolyOculus array:* The 7-pack PolyOculus module provides an aperture equivalent to that of a telescope with 0.9-meter diameter — a substantial increase, and quite large compared to typical low-cost COTS telescopes. Furthermore, the PolyOculus approach is scalable to larger apertures. In the case of the Phase I Cosmic Accelerometer, we require 28 of the 7-pack units combined into a single-output full array for sensitivity/multiplexing. This can then be sub-divided into 4x 2.5m-equivalent subarrays for multiplexing observations and stellar ensemble calibrations.

Our simulations show that the 2.5m sub-arrays should be able to reach velocity precisions of ~1 cm/s in <1-hr observations of stars as faint as V~10 mag. This will provide substantial overlap with the TESS transit catalog, for instance. Thus, the individual sub-arrays can be used to simultaneously observe multiple target stars with wide sky separations, contributing to the statistical build-up of the stellar ensemble average. At the same time, the full array can be used to study fainter stars, as well as for cross-calibration of the sub-arrays in the same manner proposed for the ultimate redshift drift experiment.

*HPRV spectrograph*: The spectrograph for the Phase I Cosmic Accelerometer will be a cross-dispersed HPRV echelle spectrograph covering 390-950nm, with the added feature of active stabilization as described in Section II.C. The spectral resolution will be $R \simeq 50,000$ for the full array and $R \simeq 100,000$ for the individual 2.5m sub-arrays. As shown in the simulations above, and established by other currently working "standard" HPRV spectrographs, these resolutions are well-suited to HPRV measurements of exoplanets orbiting low-mass stars.

*Observing strategy and expected yield:* The Phase I facility will conduct a radial velocity survey for terrestrial-mass planets in the Habitable Zone around Sun-like stars. We will follow similar strategies to current exoplanet RV surveys – looking both at TESS targets and at other stars (since transits require nearly-edge-on systems, while RV studies are more relaxed with respect to inclination angle). The survey will take 3 years to give >2 full orbits of planets in the Habitable Zone. This is particularly crucial given that stellar RV "jitter" is expected to have amplitudes ~30 cm/s [23], so that multiple orbits are expected to be necessary to allow possible identification and removal of such secular variations from the RV curve (see for instance [24]). We note that even with measurement precisions of 1 cm/s, disentangling stellar activity RV variations from orbital signatures is at the current cutting edge of the field. As such, the specific yield of this survey for exoplanets is difficult to predict. However, the demonstration of orbital RV retrieval below the stellar activity "noise floor" [24] indicates that this should be possible in some cases – and the discovery of even one Earth-like planet in the Habitable Zone of a Sun-like star would be sufficiently impactful to motivate a survey such as this.

## III. Technology Drivers

In the subsections below, we describe the key development plans and timescales for the first-phase Cosmic Accelerometer.

### A. PolyOculus Status and Development

We have previously deployed an array of four telescopes outfitted with custom low-cost acquisition and guide units with fiber optic feeds for on-sky testing. We developed an all-sky pointing model for the telescopes that meets the requirement that autonomous pointing of the telescopes brings targets within the acquisition field-of-view for full-sky pointing. We also demonstrated that individual 14-inch telescope units can guide on stars with *r*>16 mag using the low-cost uncooled Acquisition/Guide CCD. We also developed and tested the required photonic linkage, again demonstrating low cost and reliable, quality performance.

Moving to a 5m-equivalent Phase I Cosmic Accelerometer will take 5 years. During the first two years we will first demonstrate key on-sky performance parameters of a full PolyOculus array, including:

1. Fully-autonomous acquisition, guiding, and focusing while coupled to a science-grade spectrograph.

2. Multi-target spectroscopy, as needed for long-term calibration in precision radial velocity studies.

To achieve this demonstration, we will deploy a 1.6-m "Phase 0" demonstrator of the PolyOculus array. This array will be co-sited with Evryscope-North [25] at the Mount Laguna Observatory in southern California. It will also include a small, low-cost double spectrograph as its scientific detector unit with spectral resolution modes from R~200 to R~1200. This demonstrator will provide low-latency spectroscopic identification of any transient detected by Evryscope in <1 hour, and follow more numerous fainter transients detected by other transient discovery facilities down to >20 mag. It will also provide critical spectroscopic follow-up for a broad range of Evryscope, TESS, and other investigations already underway, ranging from AGN to stellar physics to exoplanets (see for instance a related Astro2020 White paper [26]). The 1.6m aperture consists of three 7-pack modules, allowing for the multi-target spectroscopy demonstration, while also providing acceptable overall sensitivity for the planned scientific observations. This effort during the first 24 months will effectively retire the risk related to the PolyOculus technology for the Phase I Cosmic Accelerometer. The estimated cost of this development is ~$800K, including hardware (array + spectrograph), personnel, deployment, and a 6-month scientific demonstration phase.

### B. Actively-stabilized spectrograph development

We have a detailed design for a prototype testbed spectrograph, which will demonstrate the key performance of the actively-stabilized spectrograph using LISA-like laser cavities and thermal actuators to stabilize the precision to < 1 cm/s. This will also allow "tuning" of the control system parameters needed for the Phase I Cosmic Accelerometer spectrograph. The key components from this can then be re-used in the actual on-sky HPRV spectrograph described in II.E above.

This development will take place in parallel with the Phase 0 demonstrator described in the preceding section.

## IV. Organization, Partnerships, and Current Status

Design and development of the Cosmic Accelerometer is led out of the University of Florida. Team members at Florida have longstanding expertise designing facility-class astronomical spectrographs as well as high-precision active laser control systems associated with LIGO and LISA. The scientific team includes some of the leading experts in the world on cosmological redshift drift experiments, as well as expertise in exoplanet research. This project is currently at the R&D level, but is ready to progress to deployment of the first 1.6m demonstrator system. The fiber-coupling technique has been demonstrated, as has the software control of the telescope array. The HPRV spectrograph is at the design stage; demonstration of the active stabilization is the next step and a pre-requisite to construction of the 5m Phase I array. Funding for this demonstration is not yet secured. No formal partnerships are in place at this time, but given existing interest in Europe in both areas of science, this facility is potentially a good candidate for joint support by US and European funding agencies.

## V. Schedule & Cost Estimates

The basic outline of our proposed schedule is as follows. We anticipate a 2-year "Phase 0" for the preliminary 1.6m demonstrator and the laboratory demonstration of the active spectrograph stabilization. This will be followed by a 3-year development and deployment period for the additional PolyOculus units to reach 5m Phase I array, as well as the full HPRV spectrograph. We then baseline a 3-year Phase I science operations period as noted above. This gives a total timeline for the proposed activities of 8 years.

For our cost estimation, we used the same costing analysis used at the University of Florida to develop bids for our fixed-price spectrograph contract bids (including FLAMINGOS-2 at >$4M, and MIRADAS at >$11M). For the Phase 0 demonstrator we estimate a total cost of $1.8M ($0.8M for the 1.6m demonstrator, and $1.0M for the spectrograph study). Extrapolating from that estimate for the PolyOculus array, and using the same costing approach for the HPRV spectrograph, we estimate construction costs at $5M for the Phase I 5m PolyOculus array, and an additional $5M for the HPRV spectrograph. With an additional $3M estimated for the 3-year operations phase (10% of the construction cost per year), we have a total cost of <$15M (including the Phase 0 demonstrator).

This Phase I facility will serve as the core for developing a future full-scale Phase 2 Cosmic Accelerometer for the redshift drift experiment. There is no technology development between Phase I and Phase 2 – the *only* difference is collecting area, which means that scaling up in aperture can be done relatively quickly. Our estimated total cost for such an experiment is $70M, including a 10-year operations budget.